\documentclass[epj]{svjour}

\usepackage{graphicx}
\usepackage{fancyhdr}

\def\mytitle{My title} 
\def\myauthors{My name}  
\def\mytype{My type of session}
\def\mysession{My session}
\def\mytitle{Higgs boson in non-minimal models} %Put your title here!
\def\myauthors{Nancy Marinelli}    %Put your name here!
\def\mytype{Contributed Talk}    
\def\mysession{Colliders - Higgs Phenomenology}
\begin{document}
\title{Higgs boson in non-minimal models at the LHC}
\subtitle{}
\author{Nancy Marinelli\inst{}
% \thanks is optional - remove next line if not needed
\thanks{\emph{Email:} nancy.marinelli@cern.ch}%
%\and
% Second author\inst{2}
% etc
% \thanks is optional - remove next line if not needed
%\thanks{\emph{Present address:} Insert the address here if needed}%
}                     % Do not remove

%
%\offprints{}          % Insert a name or remove this line
%
\institute{University of Notre Dame, Notre Dame, Indiana, USA}
%\and the second institute
%address here

%
%\date{Received: date / Revised version: date}
% The correct dates will be entered by Springer
\date{}
\abstract{
While approaching the start of the data taking at the LHC, ATLAS and
CMS perform studies involving the Higgs boson within non-minimal models besides
Supersymmetric models. Highlights from both experiments are summarized; all results 
refer  to  LHC low luminosity conditions of \mbox{10$^{33}$ cm$^{-2}$ s$^{-1}$ }.
%
%\PACS{
%      {PACS-key}{discribing text of that key}   \and
%      {PACS-key}{discribing text of that key}
%     } % end of PACS codes
} %end of abstract
\maketitle

\vspace{-3cm}
\section{Introduction}

Despite its  longstanding huge predictive power, the Standard Model 
of Electroweak interactions (SM)
%~\cite{StandardModel} 
is affected by a few flaws. 
One of them is the famous ``hierarchy problem'':  
if radiative corrections to the Higgs boson mass are computed using an 
ultra-violet cut-off $\Lambda$, the resulting value for the Higgs boson mass is 
if the order of  $\Lambda$ unless a very delicate cancellation takes place.
The most important radiative corrections to the Higgs boson mass 
arise from loops involving the top quark, gauge bosons and the Higgs itself.
An extremely un-natural fine tuning is required 
to keep the Higgs boson mass of the order of the electroweak energy scale.

Aside from  just passively accepting that Mother Nature might be so 
fine-tuned other viable theoretical solutions can be attempted,
which either stabilize the Higgs boson mass through additional symmetries and/or
shift the cut-off through some mechanism.
%to all orders and up to the Planck scale (Supersymmetry), or at the lowest order and up
%to $\sim 10$TeV  (Little Higgs models, Left-Right Symmetric Models).
Supersymmetry is nowadays the most credited solution to the hierarchy problem
and many write-ups on the subject can be found in these proceedings.

This paper is devoted instead to present studies on simulated data
devoted to investigate discovery prospects within other theoretical frameworks such as
Little Higgs, Left-Right Symmetric and Randall-Sundrum models. 
Each of the following sections will be devoted to one of these models;
a brief reminder about the phenomenological focal points if the model is followed by the
relevant studies on simulated data. 

\section{Little Higgs Model}
\label{sec1}
In the Little(st) Higgs model~\cite{LittleHiggs} the SM Higgs boson remains light because of
the introduction of a global symmetry which breaks at the TeV energy scale.
The global symmetry implies the existence of new heavy gauge bosons 
(W$^{\pm}_{\mathrm H}$,Z$_{\mathrm H}$,  $\gamma_{\mathrm H}$) and of a new heavy 
Top quark in the 1 TeV scale as well as a triplet of heavy Higgs bosons
($\Delta^{\pm\pm}$,$\Delta^{\pm}$,$\Delta^0$) in the 10 TeV scale.
They all contribute to cancel the one-loop quadratic divergences in
the Higgs boson mass. From the phenomenological point of view is
important to emphasize that a) the SM Higgs boson does still exist in the model
and the experimental standard searches apply as usual
%~\cite{SMHiggsStandardSearches}; 
b) the new Higgs bosons in the model have rather lose mass constraints. 

\subsection{Experimental expectations}
\begin{figure}[h!]
  \begin{center}
    \includegraphics[width=0.40\textwidth,height=0.40\textwidth,angle=0]{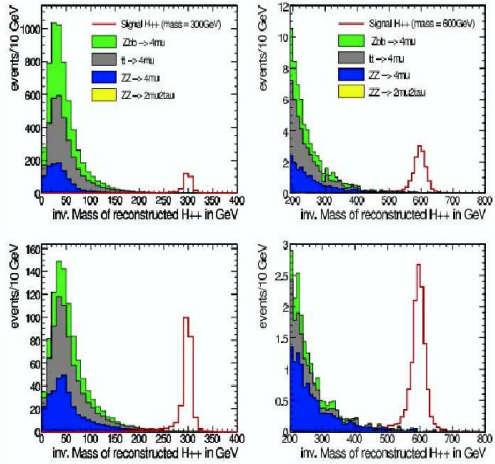}
   % \special{psfile=fig1.eps   voffset=-180 hoffset=-90 hscale=60 vscale=60 angle=0}
  \end{center}
  \vspace{-0.6cm}
  \caption{Reconstructed same-sign muon invariant mass after online selection (upper plots)
and after offline selection (lower plots). Plots correspond to an integrated luminosity of
10 fb$^{-1}$. }
  \label{fig1}      
 \end{figure}

\begin{figure}[h!]
  \begin{center}
    \vspace{5cm}
    \includegraphics{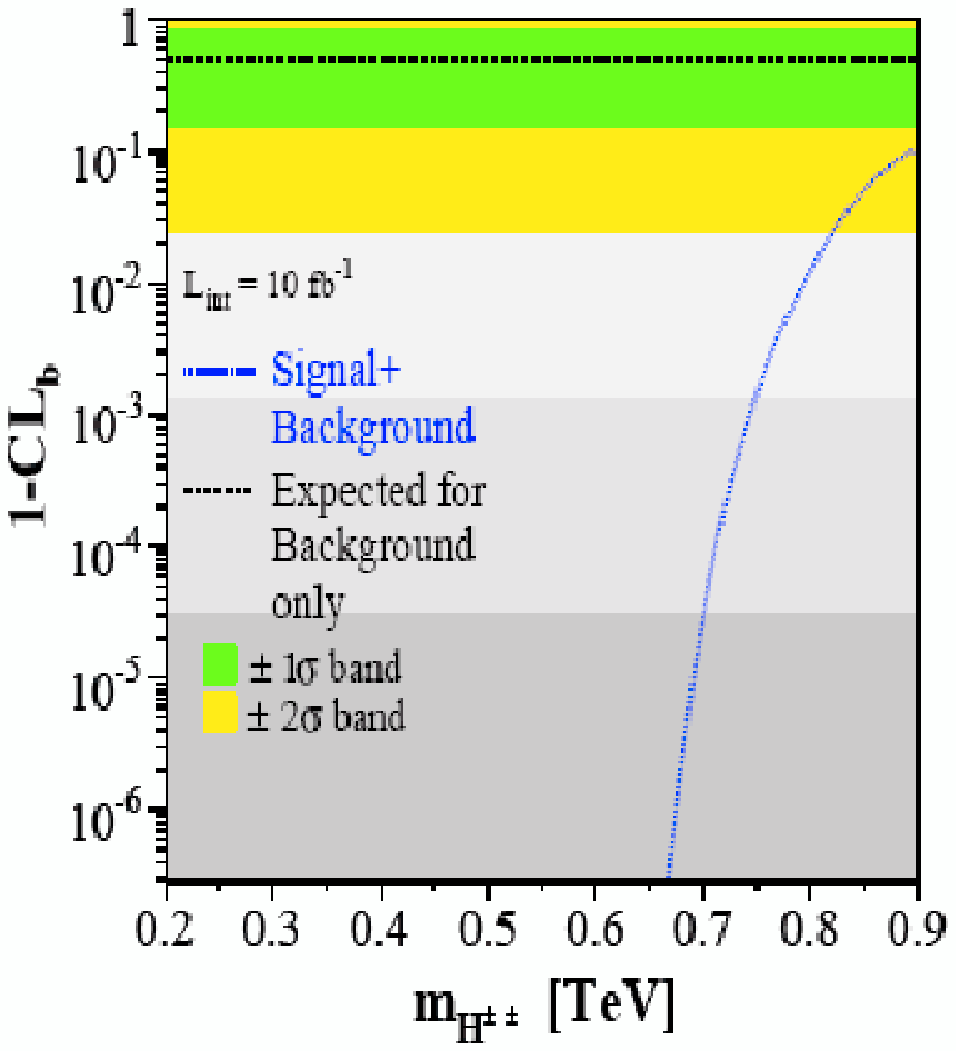}
    \includegraphics{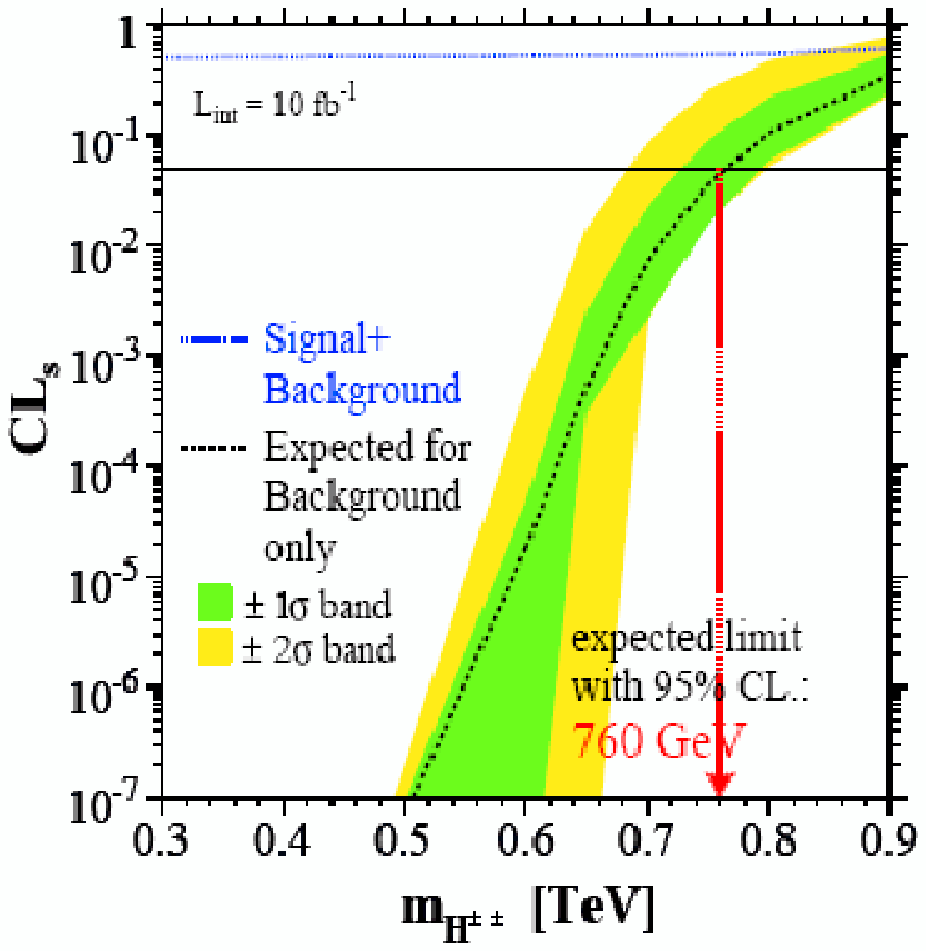}
    
  \end{center}
  \caption{Left: 1-CL$_b$ as a function of the reconstructed  $\Delta^{\pm\pm}$ mass: the discovery limit
is 650 GeV. Right: CL$_s$ as function of  the reconstructed  $\Delta^{\pm\pm}$ mass: the exclusion limit is
760 GeV. Results are shown for an integrated luminosity of 10 fb$^{-1}$. }
  \label{fig2}     
  
\end{figure}

The CMS Collaboration has investigated, using full detector simulation, 
 the discovery potential of doubly-charged Higgs bosons in the Drell-Yan pair-production process
(${\mathrm q}\overline{\mathrm q} \rightarrow \Delta^{++}\Delta^{--}$) 
with the $\Delta^{\pm\pm}$ decaying to same-sign muon pairs~\cite{cms_DoublyChargedHiggs}.
The decay branching fraction to muon pairs was assumed to be 100\%. 
The four-muon final state provides an exceptionally clear signature with 
SM background naturally small. The muon pair invariant mass shows the signal 
already after the online event selection and only fewer additional loose cuts enhance it
to the level shown in Fig.~\ref{fig1} where distributions are shown for an integrated 
luminosity of 10 fb$^{-1}$.  
The statistical interpretation of the invariant mass spectrum leads to a discovery 
limit of 650 GeV and  exclusion up to 760 GeV (Fig.~\ref{fig2}).
The search of doubly-charged Higgs bosons was also investigated by ATLAS in the
single production via Vector Boson Fusion~\cite{atlas_VBFDoublyChargedHiggs}. 
The results, however, were not very encouraging since the sensitivity was poor
even with an integrated luminosity of 300 fb%{-1}$.

\section{Left-Right Symmetric Models}
\label{sec2}
In Left-Right Symmetric Models (LRSM) two (left and right handed) Higgs
boson triplets provide a parity-conserving Lagrangian so that 
all fermions are treated symmetrically as Left Handed and
Right Handed doublets~\cite{LRSM}.

Based on SU(2)$_{\mathrm L}\times$SU(2)$_{\mathrm R}\times$U(1)$_{\mathrm {B-L}}$, 
the new symmetry 
leads to the introduction of new neutrinos, bosons and Higgs bosons,
in particular doubly-charged Higgses as it was the case in the Little Higgs
model.
A peculiar aspect of the LRSM is that the Yukawa couplings of the Higgs
boson triplet allow the existence of Majorana mass terms of the Right-Handed neutrinos,
rendering the see-saw mechanism possible~\cite{see_saw}. 
The LRSM is hence capable to solve both the hierarchy problem and
to naturally explain the experimental evidence of low, 
non-zero mass of the Left-Handed neutrinos~\cite{neutrinoMass}.
\subsection{Experimental expectations}
\begin{figure}[h!]
  \begin{center}
    \vspace{0.5cm}
    \includegraphics[width=0.30\textwidth,height=0.35\textwidth,angle=0]{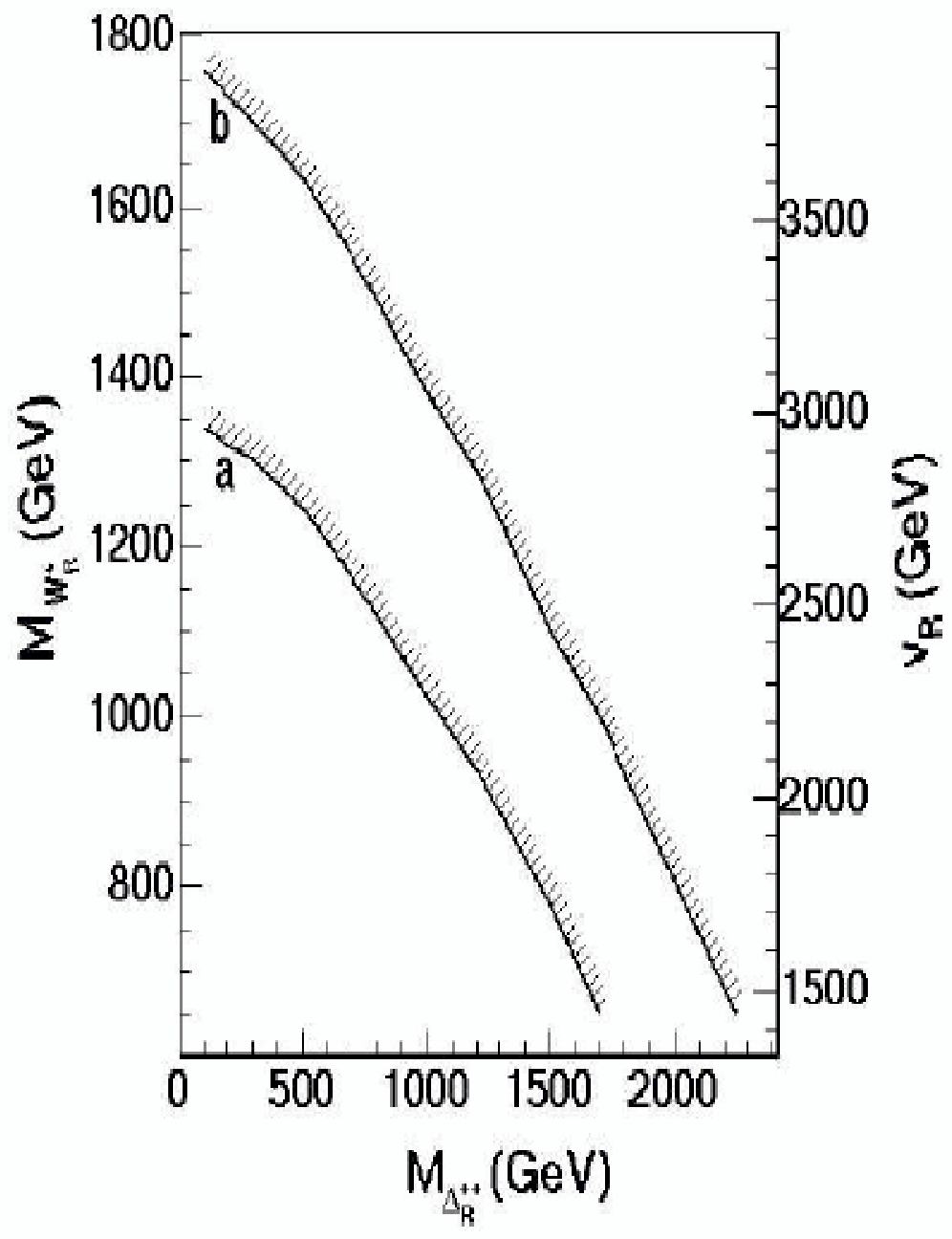}
  \end{center}
  \vspace{-0.6cm}
  \caption{Discovery reach for \mbox{$\Delta^{++}_R\rightarrow l^+l^-$} in the \mbox{ (m$_{\Delta^{++}_R}$, m$_{W_R}^{+}$) } plane
    for integrated luminosity of 100 fb$^{-1}$ (a) and 300 fb$^{-1}$ (b).}   
  \label{fig3}      
\end{figure}

The LRSM has a number of interesting signatures. The interest however is
here concentrated on Higgs bosons. As in the case of the Little Higgs,
the observation of doubly-charged Higgs would provide important evidence
of new physics. Search techniques of left-handed doubly-charged Higgs
go along the lines of what already seen in Sec.~\ref{sec1}. 
The ATLAS collaboration has also probed the discovery reach for right-handed doubly-charged 
Higgs~\cite{atlas_DoublyChargedHiggs}, in single production via VBF with decay into lepton
pairs (e,$\mu$). Since the background is negligible, discovery can be claimed if the number 
of signal events is greater than 10. The discovery reach is illustrated in Fig.~\ref{fig3}.

\section{ExtraDimensions: Randall-Sundrum model}
\label{sec3}
In its basic formulation~\cite{RandallSundrum} the Randall-Sundrum model introduces 
only one extra dimension so that the Universe is described in terms of two four-dimensional Minkowskian branes
bounding a five-dimensional bulk. The so called ``weak gravity'' brane is the one
where the world we know sits, i.e. where the ElectroWeak and TeV scale processes occur.
At the other end of the five-dimensional space, separated by a distance
proportional to the ``compactification radius'' r$_c$, 
there is instead the so called ``strong gravity'' brane, where gravity is at the Plank scale. 
The law describing the relation between the two branes is a decreasing exponential governed by the    
``warp factor'' kr$_c$. The hierarchy between the Planck and ElectroWeak scale is removed 
if kr$_c\sim$ 12 so that the $\,$ compactification $\;$ radius $\;$ is extremely small 
\mbox{($\sim10^{-32}$m)}  and no deviations  from Newton's law are visible at experimental level. 
The RS model makes a number of predictions, the existence of the graviton being the central one.
However more interesting in the context of this paper is the fact that to stabilize the size of
the extra dimension (i.e.  kr$_c\sim$ 12 ) it was necessary, after the first formulation of the model,
to introduce the radion  $\Phi$~\cite{GoldbergerWise}, 
representing  the fluctuations of the distance between the two branes.
The RS scalar sector has only four free parameters 
(
m$_\Phi$= radion mass, 
m$_{\mathrm h}$= SM Higgs mass, 
$\Lambda_\Phi$= radion v.e.v.  and $\xi$= $\Phi$-h mixing).
Very important aspects are that $\Lambda_\Phi\sim$1TeV and
m$_\Phi<$1 TeV without need of fine tuning, the radion and the SM Higgs boson couplings to 
gauge bosons and fermions are very similar and, finally, that the radion and the Higgs boson
mix.  The latter point implies that for certain regions of the parameter space
the decay channel $\Phi\rightarrow$hh opens up and can be investigated in the classical Higgs boson
decay channels, i.e.  
$\gamma\gamma {\mathrm b}\overline{\mathrm b}$, $\tau\tau {\mathrm b}\overline{\mathrm b}$
and ${\mathrm b}\overline{\mathrm b}{\mathrm b}\overline{\mathrm b}$. 

\subsection{Experimental expectations}
ATLAS and CMS have both concentrated their work on the 
$\gamma\gamma {\mathrm b}\overline{\mathrm b}$ and $\tau\tau {\mathrm b}\overline{\mathrm b}$
final states~\cite{atlas_radion,cms_radion}, ${\mathrm b}\overline{\mathrm b}{\mathrm b}\overline{\mathrm b}$ being very difficult
and almost hopeless because of the huge multi--jet background which affects it.
ATLAS and CMS adopt two different approaches for the search in $\gamma\gamma {\mathrm b}\overline{\mathrm b}$;
while ATLAS assumes that the SM Higgs has already been discovered and its mass measured, 
CMS has developed an analysis strategy
aiming to discovering the radion and the Higgs boson at the same time.

The general event selection is based on requiring two high-p$_T$, isolated photons and two
high-p$_T$ jets of which at least one must come from a b-quark.
Backgrounds to this channel come from 
${\gamma\gamma \mathrm{b\overline{b}}}$, 
${\gamma\gamma \mathrm{bj}}$, 
${\gamma\gamma \mathrm{jj}}$, 
${\gamma\gamma \mathrm{cj}}$ and
${\gamma\gamma \mathrm{c\overline{c}}}$ but they are on overall small.
PYTHIA and fast detector simulation were used by ATLAS for both signal and background; 
PYTHIA and full detector simulation were used by CMS to produce  signal samples while
MadGraph~\cite{MadGraph}, CompHep~\cite{CompHep} and fast detector simulation were used for the
backgrounds.

\begin{figure}[ht!]
  \begin{center}
    \vspace{-0.4cm}
    \includegraphics[width=0.50\textwidth,height=0.35\textwidth,angle=0]{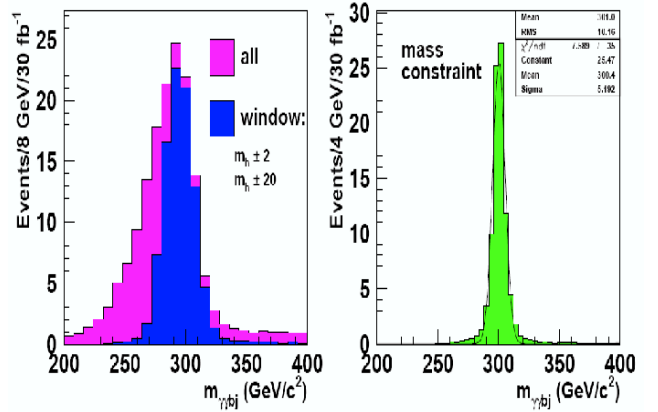}
  \end{center}
  \vspace{-0.4cm}
  \caption{Reconstructed $\gamma\gamma \mathrm{bj}$ invariant mass distribution for
    m$_{\mathrm h}$=125 GeV,  m$_\Phi$=300 GeV, $\xi=0$, $\Lambda_\Phi= 1$ TeV and for 30 fb$^{-1}$. The plot on 
    the left shows all combinations as well as those fulfilling the mass window cuts. 
    The  plot on the right is obtained by constraining the reconstructed masses m$_{\mathrm {bj} }$
    and  m$_{\gamma\gamma}$ to the light Higgs mass m$_{\mathrm h}$, after the mass window cuts.
  } 
  \label{fig4}      
\end{figure}

In ATLAS the di-photon and di-jet invariant masses were $\;$ calculated and
mass $\;$ window cuts applied: \mbox{m$_{\gamma\gamma}$=m$_{\mathrm h}\pm 2$ GeV} and 
m$_{\mathrm {bj} }$=m$_{\mathrm h}\pm 20$ GeV. Photons and jets satisfying these
conditions were combined to calculate the m$_{\gamma\gamma \mathrm{bj}}$ invariant mass
as shown in Fig.~\ref{fig4}.  

A constraint of the Higgs boson invariant mass to its
``known'' value improves the resolution.
Since the background in this channel is low, a discovery can be claimed if
at least ten signal events are found. An integrated luminosity of
30 fb$^{-1}$ leads to a reach in $\Lambda_\phi$ up to 2.2 TeV for
m$_\Phi$=300 GeV and up to 0.6 GeV for  m$_\Phi$=600 GeV.

\begin{figure}[h!]
  \begin{center}
    \includegraphics[width=0.50\textwidth,height=0.30\textwidth,angle=0]{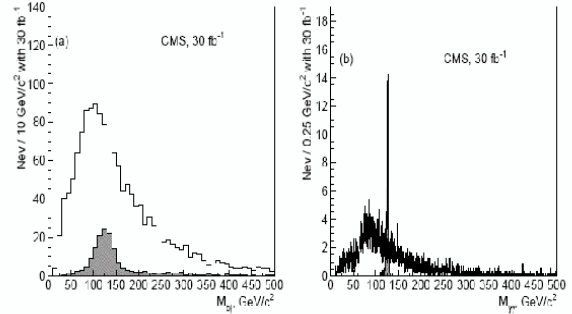}
  \end{center}
  \vspace{-0.6cm}
  \caption{(a) Di-jet invariant mass and (b) di-photon invariant mass. The signal is shown by the solid histograms.
The distributions were obtained after all selection criteria but the mass window had been applied. The signal is shown
for the more favourable point in the ($\xi,\Lambda_\Phi$) plane.}
  \label{fig5}      
\end{figure}

According to its different approach, CMS would attempt a simultaneous discovery of the Higgs boson
and of the radion. The observation of a peak in the di-photon mass distribution  obtained from the selected
sample of ${\gamma\gamma \mathrm{bj}}$ events would indicate the presence of one of the two Higgs bosons (Fig.~\ref{fig5}). 
Di-photon events falling in a window of 4 GeV around the peak and di-jets events falling within
60 GeV from the $\gamma\gamma$ peak can be considered to isolate the radion signal.

\begin{figure}[h!]
  \begin{center}
    \includegraphics[width=0.40\textwidth,height=0.35\textwidth,angle=0]{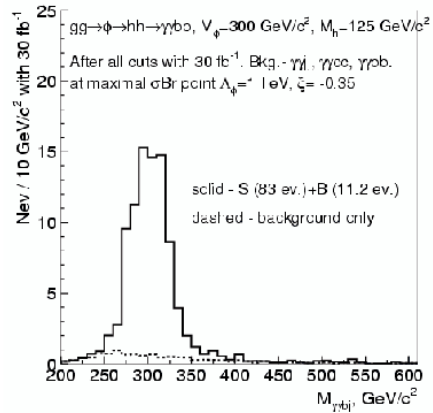}
  \end{center}
  \vspace{-0.6cm}
  \caption{${\gamma\gamma \mathrm{bj}}$ invariant mass after all selection requirements, including the mass windows for signal (solid line)
 and background (dashed line).}
  \label{fig6}      
\end{figure}

The radion can be discovered from the excess of events over the background level expected after
the same set of selection requirements have been applied (Fig.~\ref{fig6}).
The statistical interpretation of CMS discovery strategy leads to a reach in $\Lambda_\Phi$ up
to $\sim2.5$ TeV.

\begin{figure}[h!]
  \begin{center}
    \includegraphics[width=0.30\textwidth,height=0.30\textwidth,angle=0]{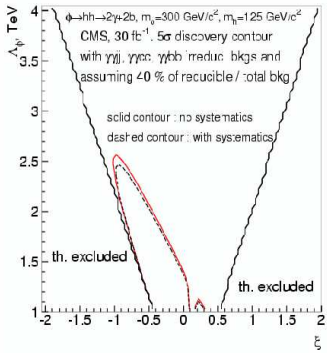}
  \end{center}
  \vspace{-0.6cm}
  \caption{The 5-$\sigma$ contour for $\Phi\rightarrow\gamma\gamma{\mathrm b}\overline{\mathrm b}$ for
  m$_{\mathrm h}$=125 GeV,  m$_\Phi$=300 GeV, $\xi=-0.35$, $\Lambda_\Phi= 1$ TeV and for 30 fb$^{-1}$.  The dashed
and solid contour lines refer to the discovery region accessible with and without including the systematic uncertainties
arising from background evaluation and from renormalization and factorization scales.}
  \label{fig7}      
\end{figure}

The study of the $\tau\tau {\mathrm b}\overline{\mathrm b}$ final state, with one
$\tau$ decaying leptonically and the other hadronically, was also performed by
both ATLAS~\cite{atlas_radion}  and CMS~\cite{cms_radion}, assuming that the Higgs mass is already known. 
%Events are generally selected online by requiring a high-P$_T$ lepton associated to 
%a $\tau$-jet. In the offline step the $\tau\tau$ invariant mass is calculated by
%combining the lepton and the  $\tau$-jet, finally the requirement of the two jets with large
%transverse energy, of which at least one must cone from a b-quark,  is added.
Results obtained with this channel, however, are not so encouraging as for the $\gamma\gamma {\mathrm b}\overline{\mathrm b}$ channel.
In general the signal efficiency is low and the analysis risks to be largely dominated
by systematic errors due to the evaluation of the background. The expected reach in $\Lambda_\Phi$ is up to $\sim$1 TeV.

\section{Conclusions}
Non-minimal models such as those briefly presented in this paper, are becoming
popular as alternative solutions to the hierarchy problem which afflicts the Standard Model.    
Both ATLAS and CMS Collaborations are investigating the discovery potential 
of the new ``zoo'' of particles necessarily introduced by the models to cancel
the divergent radiative corrections to the Higgs boson mass. 

The search of doubly-charged Higgs bosons was investigated both in VBF and Drell-Yan
production. The sensitivity at large masses ($>$ 1 TeV) seems to be rather poor, however
discovery can be achieved up to $\sim$ 650 GeV (exclusion up to $\sim$ 750 GeV).

Right Handed doubly-charged Higgs bosons appearing in the LRSM  can be probed
in the purely leptonic channel up to $\sim$ 1.7 TeV (100 fb$^{-1}$). 

Finally, it should be possible to probe the Randall-Sundrum scalar sector 
up to $\Lambda_\Phi\sim$ 2.5 TeV by looking at $\Phi\rightarrow {\mathrm {hh}}\rightarrow \gamma\gamma{\mathrm b}\overline{\mathrm b}$.


\begin{thebibliography}{999}
%
% and use \bibitem to create references.
%

%\bibitem{StandardModel}


%\bibitem{SMHiggsStandardSearches}

\bibitem{LittleHiggs}
N. Arkani, A.G. Cohen, E. Katz and A.E. Nelson, JHEP, \textbf{0207}, (2002) 034;
N. Arkani, A.G. Cohen and H. Georgi, Phys. Lett. \textbf{B513}, (2001) 232-240; 
T. Han et. al, Phys. Rev. \textbf{D67} (2003) 095004.

\bibitem{LRSM}
R.N. Mohapatra and J.C. Pati, Phys. Rev. \textbf{D11} (1975) 566;
R.N. Mohapatra and J.C. Pati, Phys. Rev. \textbf{D11} (1975) 2558.

\bibitem{see_saw}
R.N. Mohapatra ans P.B. Pal, Phys. Rev. \textbf{D38} (1988) 2226;
R.N. Mohapatra, arXiv:hep-ph/0412379v1;



\bibitem{cms_DoublyChargedHiggs}
T.Rommerskirchen and T.Hebbeker, \textbf{CMS Note 2006/081}.

\bibitem{atlas_VBFDoublyChargedHiggs}
G. Azuelos et al., Eur. Phys. J. \textbf{C3962}, (2005), 13-24. 


\bibitem{neutrinoMass}
Y. Fukuda et al., Phys. Rev. Lett. \textbf{81}, (1998) 1562;
Y. Fukuda et al., Phys. Rev. Lett. \textbf{B539}, (2002) 179.


\bibitem{atlas_DoublyChargedHiggs}
G. Azuelos et al., J. Phys.  \textbf{G32}, (2006), 73-92.

\bibitem{RandallSundrum}
L. Randall and R. Sundrum, Phys. Rev. Lett. \textbf{83}, (1999) 3370.

\bibitem{GoldbergerWise}
W.D. Goldberger and M. B. Wise, Phys. Rev. Lett. \textbf{83}, (1999) 4922.


\bibitem{atlas_radion}
G. Azuelos et al., Eur. Phys. J. direct \textbf{C4}, (2002) 16.


\bibitem{cms_radion}
D. Dominici et al., \textbf{CMS Note 2005/007}.

\bibitem{MadGraph}
F. Maltoni and T. Stelzer, Journ. of Hight Energy Phys., \textbf{0302}, (2003), 027;
T. Stelzer and W.F. Long, Comput. Phys. Commun., \textbf{81}, (1994) 357.

\bibitem{CompHep}
A. Puckhov et al., hep-ph/9908288.

% Format for books
%\bibitem{RefB}
%Author, \textit{Book title} (Publisher, place year) page numbers
% etc


\end{thebibliography}
\end{document}